\documentclass[12pt]{article}
\usepackage{pslatex}
\usepackage{amssymb}
\usepackage{graphicx}
\usepackage{color}
\usepackage{epstopdf}

\makeatletter

\@addtoreset{equation}{section}
\makeatother

\newcommand{\be}{\begin{eqnarray}}
\newcommand{\ee}{\end{eqnarray}}
\newcommand{\nn}{\nonumber}
\newcommand{\benu}{\begin{enumerate}}
\newcommand{\eenu}{\end{enumerate}}

\def\IC{\mathbb{C}}

\def\IR{\mathbb{R}}
\def\IZ{\mathbb{Z}}

\def\CI{{\cal I}}

\def\CL{{\cal L}}

\def\CN{{\cal N}}

\def\a{\alpha}
\def\b{\beta}

\def\d{\delta}

\def\th{\theta}

\def\m{\mu}
\def\n{\nu}

\def\s{\sigma}

\def\t{\tau}

\def\w{\omega}

\def\D{\Delta}

\def\S{\Sigma}

\def\O{\Omega}

\def\half{\frac{1}{2}}
\def\thalf{{\textstyle \frac{1}{2}}}
\def\imp{\Longrightarrow}
\def\iff{\Longleftrightarrow}
\def\goto{\rightarrow}

\newcommand{\bra}[1]{\langle{#1}|}
\newcommand{\ket}[1]{|{#1}\rangle}

\def\Tr{{\rm Tr}}

\def\vol{\mbox{vol}}
\def\Vol{\mbox{Vol}}
\def\bra{\langle}
\def\ket{\rangle}

\def\tw{\tilde{w}}

\begin{document}

\begin{titlepage}

\begin{flushright}
SNUST 0601-01\\
hep-th/0601223
\end{flushright}
\vspace*{2.0cm}
\centerline{\Large\bf Comments on Anomalies and
Charges of Toric-Quiver Duals}
\vspace*{0.5cm}
\centerline{\Large\bf} \vspace*{1.5cm}
\centerline{Sangmin Lee and Soo-Jong Rey}
\vspace*{1.0cm}
\centerline{\sl School of Physics \& Center for Theoretical Physics}
\centerline{\sl Seoul National University}
\centerline{\sl Seoul 151-747 KOREA}
\vskip0.3cm \vspace*{2.0cm}
\centerline{\bf ABSTRACT}
\vspace*{1cm}
\noindent We obtain a simple
expression for the triangle `t Hooft anomalies in quiver gauge
theories that are dual to toric Sasaki-Einstein manifolds. We
utilize the result and simplify considerably the proof concerning
the equivalence of $a$-maximization and $Z$-minimization.
We also resolve the ambiguity in defining the flavor charges
in quiver gauge theories.
We then compare coefficients of the triangle anomalies with
coefficients of the current-current correlators and find
perfect agreement.


\end{titlepage}


\section{Introduction}

In recent years, a large number of new examples of AdS$_5$/CFT$_4$
correspondence \cite{adscft} have been constructed and studied
extensively. IIB string theory on AdS$_5\times Y$ preserves ${\cal
N}=2$ supersymmetry (8 supercharges) when $Y$ is a Sasaki-Einstein
(SE) manifold \cite{kw, gub, ach, mp}. Soon after the discovery of
new SE metrics \cite{gmsw1, clpp}, it was realized that many of the
SE manifolds are toric \cite{ms1, msy, ms2}. When $Y$ is toric, most
geometric quantities such as its volume can be computed without
knowledge of the explicit metric \cite{msy}. The toric description
also helped identifying ${\cal N}=1$ superconformal gauge theory
duals \cite{bfhms, bk, fhm, bfz}, the quiver gauge theories. Using
new techniques to analyze quiver gauge theories, very detailed
checks have been made for toric-quiver dual pairs
\cite{fhh}-\cite{cep}.

One such issue concerns identifying the correct $R$-symmetry at the
conformal fixed point. The superconformal $U(1)_R$ symmetry is in
general a nontrivial linear combination of all nonanomalous global
$U(1)$ symmetries. In gauge theory dual, it was found in \cite{amax}
that maximizing $a$-function determines uniquely the correct
combination. Denoting the global charges as $Q_I$, the definition of
$a$ as a function of the trial $R$-charge contains the triangle `t
Hooft anomaly, whose coefficient is given by \footnote{Throughout
this paper, we work in the usual large $N$ limit and suppress the
dependence on $N$. It can be easily reinstated so that $a$ is
proportional to $N^2$, $F_i^I$ is proportional to $N$, etc.}
\be C_{IJK} = \Tr(Q_I Q_J Q_K). \label{c} \ee
The rule of $a$-maximiation in ${\cal N}=1$ supersymmetric gauge
theory and its geometric dual have played a crucial role throughout
the development \cite{bbc}-\cite{bz2}. 
The conserved currents $J_I$ associated with the
charges $Q_I$ are mapped to $U(1)$ gauge fields $A^I$ in
supergravity via AdS/CFT correspondence. Then the anomaly
coefficient $C_{IJK}$ is encoded \cite{witten} as the coefficient of
the Chern-Simons term in the five-dimensional gauged supergravity
action
\be S_{CS} \sim \int C_{IJK} \, A^I \wedge F^J \wedge F^K. \ee
The anomaly coefficients $C_{IJK}$ is also suggested intimately
related to the coefficients $\t_{IJ}$ of the two-point correlators
among conserved currents via $\t_{RR}$ minimization \cite{trr}.

While the gauge theory expression for $C_{IJK}$ (\ref{c}) is now
available from \cite{bk,bz}, the supergravity expression in terms of
geometric data on SE manifold has been lacking so far (see, however,
the paragraphs below). On the contrary, the expression for $\t_{IJ}$
is known in supergravity \cite{bgiw} but not in the gauge theory. To
make a connection between $C_{IJK}$ and $\t_{IJ}$ as suggested in
\cite{trr}, one thus needs a more geometric understanding of
$C_{IJK}$. In fact, from the supergravity viewpoint, the connection
ought to exist since $\t_{IJ}$ and $C_{IJK}$ are both derivable from
an underlying prepotential ${\cal F}$ \cite{gaugedsugra}.

In this work, we report progress in comparing global charges and
anomalies from gauge theory and those from supergravity. In
particular, we identify the flavor charges in gauge theory
unambiguously and use the identification to compare the expression
for triangle 't Hooft anomalies in supergravity and gauge theory.

Our work begins in section 2 with a simple observation that the
gauge theory result for the triangle `t Hooft anomaly coefficients
as derived in \cite{bk, bz} is nothing but the area of a triangle 
connecting three vertices on the toric diagram:
\be \label{cc} C_{IJK} = \frac{1}{2} |\bra v_I,
v_J, v_K \ket|. \ee
After deriving this formula, we illustrate its use by re-deriving
the equivalence \cite{bz} of $a$-maximization and its geometric
counterpart, $Z$-minimization \cite{msy}. Although our proof is
similar to the original one \cite{bz}, the use of (\ref{cc}) reduces
the amount of needed computation considerably. We also resolve
the ambiguity in defining the non-$R$ `flavor' charges in the gauge
theory so as to facilitate the comparison with supergravity results.

Clearly, the next logical step is to compute $C_{IJK}$ in
supergravity by performing perturbative Kaluza-Klein (KK) reduction
up to cubic order. While we were making progress in that direction,
Ref. \cite{bpt} appeared, in which a supergravity formula for
$C_{IJK}$ valid for any (not necessarily toric or Sasakian) Einstein
manifold, as well as the gauge theory result (\ref{cc}), were
obtained. Section 3 of our paper is organized accordingly. After
reviewing the linearized approximation to KK reduction and fixing
the normalization of the charges, we show that the flavor charges
computed in field theory in section 2 agrees perfectly with the
supergravity result \cite{bgiw}. Finally, we make an explicit check
of the relation $\t_{IJ} = - 3 C_{RIJ}$ \cite{trr} using the result
from \cite{bgiw, bpt} and again find perfect agreement.

\section{Toric quiver gauge theory side}
It is by now well-known that the global $U(1)$ symmetries of a gauge
theory with an SE dual are divided into two kinds. One is called
baryon symmetry, and corresponds to $D3$-branes wrapping calibrated
three-cycles of the SE manifold $Y$. The other is often called
flavor symmetry and is associated with the isometry of $Y$. How the
gauge fields for each $U(1)$ symmetry arise in the ${\rm AdS}_5$
gauged supergravity will be reviewed in section 3.

In the toric case, $Y$ has three isometries by definition, and the
number of independent three-cycles are given by the toric data. Both
symmetries are most efficiently described in the language of toric
geometry, not only on the supergravity side but also in the quiver
gauge theory. So, we shall begin with a quick review of well-known
facts about the toric geometry of $Y$,  mainly to establish our
notations and summarize some results pertinent to discussion in
later sections. See \cite{ms1, msy} for more information on toric
geometry in this context.

\subsection{A short review of toric SE manifolds \label{toreview}}

It is useful to define the SE manifold $Y$ in terms of the cone $X =
C(Y)$ with the metric
\be
\label{cone}
ds^2_{X} = dr^2 + r^2 ds^2_{Y} .
\ee
The manifold $Y$ being Sasakian is equivalent to the cone $X$ being K\"ahler.
The Reeb Killing vector field defined as
\be
K_R = \CI \left( r\frac{\partial}{\partial r} \right),
\ee
where $I$ denotes the complex structure on $X$, is translated to the
$R$-symmetry of the field theory dual. The manifold $Y$ is
Sasaki-Einstein if $X$ is K\"ahler and Ricci-flat, i.e., Calabi-Yau
(CY). It is known that when $Y$ is SE, it can be locally described
as the $U(1)_R$ fibration over a K\"ahler-Einstein base $B$. The
following relations will be useful when we prove some identities in
section 3: \footnote{Generically, $B$ is an orbifold rather than a
smooth manifold. Some of the proofs in section 3 involve integration
by parts over $B$, hence they are not strictly valid. But, we expect
that similar proofs will work with mild modifications.}
\be ds_X^2 = dr^2 + r^2((e^0)^2 + ds_B^2), && e^0 \equiv \frac{1}{3}
d\psi + \s, \qquad K_R = 3 \frac{\partial}{\partial \psi},
\nn
\\
\label{relations} J_X = r^2 J_B + r dr \wedge e^0 , && \O_X =
e^{i\psi} r^2 \O_B \wedge (dr +ir e^0),
\\
R_{\m\n}^{(B)} = 6 g_{\m\n}^{(B)}, && d\s = 2 J_B, \qquad d\O_B = 3
i \s \wedge \O_B.
\nn
\ee

In physics terminology, a toric cone $X$ is conveniently described
by the gauged linear sigma model (GLSM). For $X$, the GLSM takes a
D-term K\"ahler quotient of $\{ Z^I \} \in \IC^{d}$ with respect to
integer charges $Q_a^I$:
\be
\label{glsm}
\sum_{I=1}^d Q_a^I |Z^I|^2 =0,
\;\;\;\;\;
Z^I \sim e^{i Q_a^I \th^a} Z^I
\;\;\;\;\;
(a = 1, \cdots, d-3) ,
\ee
leaving a three-dimensional complex cone.
The CY condition
sets $\sum_I Q_a^I = 0$ for each $a$.

Let $\{v^i\}$ ($i=1,2,3$) be a basis of the kernel of the map $Q_a :
\IZ^d \goto \IZ^{d-3}$, i.e.,  $Q_a^I v_I^i = 0$. One can regard
$v_I^i$ as $d$ lattice vectors in $\IZ^3$ and use them to
parameterize $|Z^I|^2 = v_I \cdot y \equiv v_I^i y_i$ $(y \in
\IR^3)$. The allowed values of $y$ form a polyhedral cone $\D$
defined by $\{ v_I \cdot y \ge 0\}$ in $\IR^3$. The cone $X$ is then
a fibration of three angles $\{\phi^i\}$ over the base $\D$. Using
the CY condition $\sum_I Q_a^I  = 0$, one can set $v_I^1=1$ for all
$I$, as this assignment satisfies $Q_a^I v_I^1 = 0$ automatically.
We will always set $v_I^1=1$. 
The polygon drawn on $\IR^2$ with 
the remaining components of $v_I$'s 
is usually called the toric diagram.

By definition, a toric $Y$ has three isometries $K_i =
\partial/\partial \phi^i$. The Reeb vector $K_R$ is in general a
linear combination of them, $K_R = b^i K_i$. In \cite{msy}, it was
shown that the Reeb vector characterizes all the essential geometric
properties of $Y$. The manifold $Y$ is embedded in $X$ as $Y = X
\cap \{b\cdot y = 1/2\}$. Supersymmetric cycles of $Y$ are given by
$\S^I = Y \cap \{v_I \cdot y=0 \}$.

The Reeb vector also determines a unique Sasakian metric on $Y$. 
The volume of $Y$ is computable by summing over the volume of the
supersymmetric cycles \cite{msy}:
\be \Vol(Y) = \frac{\pi^3}{b^1} \sum_I \frac{\langle
v_{I-1},v_I,v_{I+1} \rangle}{\langle b,v_{I-1},v_I \rangle \langle
b, v_I, v_{I+1} \rangle}. \ee
Here, $\langle u, v, w \rangle$ denotes the determinant of the
$(3\times 3)$ matrix made out of vectors $u, v, w$. The CY condition
on $X$ fixes $b^1 =3$. The metric of $Y$ becomes Einstein at the
minimum of $\Vol(Y)$ as $b^2, b^3$ are varied inside the polyhedral
cone: $b \in \D$.

As explained in \cite{fhm}, when $Y$ is simply-connected, which we
assume for the rest of this paper, the homology group of $Y$ is
given by $H_3(Y,\IZ)=\IZ^{d-3}$. If $\{C^a\}$ ($a=1,\cdots ,d-3$) form a
basis of three-cycles of $Y$, it can be shown that $\S^I = Q_a^I
C^a$, where $Q_a^I$ is precisely the GLSM data (\ref{glsm}) of $Y$.
The harmonic three-forms $\w_a$ dual to $C^a$ measure the baryon
charges of $\S^I$, so
\be \label{baryoncharge}
B_a\left[\S^I\right] = \int_{\S^I} \w_a =
Q_a^I . \ee

As one can see from the torus action in the GLSM description
(\ref{glsm}), the baryon charges $Q_a^I$ and the flavor charges
$F_i^I$ together span $\IZ^d$ (for simply connected $Y$). This means
that the toric relation $Q_a^I v_I^i=0$ can be extended to
\be
\label{extor}
\pmatrix{ Q_a{}^I \cr F_i{}^I } \pmatrix{ u_I{}^{b} & v_I{}^j}
= \pmatrix{ \d_a^b & 0 \cr 0 & \d_i^j },
\ee
for some integer-valued matrices $F_i^I$ and $u_I^b$.
One may want to interpret $F_i^I$ as the $i$-th flavor charges of $\S^I$,
i.e., $F_i\left[\S^I\right] = F_i^I$. However,
even after choosing a fixed basis for $v_I^i$,
the relation (\ref{extor}) does not fix $F_i^I$ uniquely,
as one may shift $F_i^I$ and $u_I^b$ by
\be
F_i{}^I \goto F_i{}^I + N_i{}^a Q_a{}^I,
\;\;\;\;\;
u_I{}^b \goto u_I{}^b - v_I{}^i N_i{}^b.
\ee
This freedom is called the mixing ambiguity in the literature;
flavor symmetry is unique up to mixing with baryon symmetries. This
immediately poses a question: in comparing the gauge theory results
with the supergravity results, how are the flavor charges on both
sides to be identified? Later in this section, we will show that
there is a unique, preferred choice of (non-integer) $F_i^I$ which
matches with the supergravity result.

\subsection{Triangle anomaly from triangle area}

We shall now derive a formula for the triangle 't Hooft anomaly of
quiver gauge theories dual to $Y$. The formula states that the
anomaly coefficient $C_{IJK} = \Tr(Q_I Q_J Q_K)$ is simply the area
of the triangle connecting the three vertices $v_{I,J,K}$ on the
toric diagram: \be \label{ccc} C_{IJK} = \frac{1}{2} |\bra v_I, v_J,
v_K \ket|. \ee The derivation of (\ref{ccc}) is built upon some
known features of the quiver gauge theories \cite{bz}:
\benu

\item
The number of gauge group $F$ is twice the area of the toric diagram.

\item
Let $w_I \equiv (v_{I+1} - v_I)$ denote the edges of the toric
diagram. Associated with each pair of edges $(w_I, w_J)$, there are
bifundamental chiral superfields $\Phi_{IJ}^r$ with the same charges
(see below) and multiplicity given by $|\bra w_I, w_J \ket | \equiv
|w_I^2 w_J^3 - w_I^3 w_J^2 |$.

\eenu See \cite{bz} and references therein for more details. The
formula (\ref{ccc}) is then derivable from the expression for the
$a$-function for the quiver gauge theories.

\begin{figure}[htbp]
   \label{cijk}
   \centering
   \includegraphics[width=7cm]{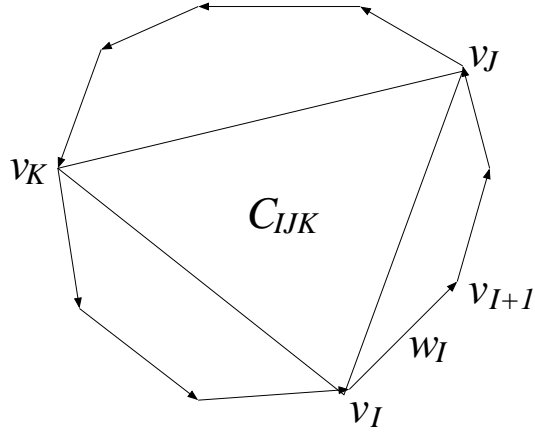}
   \caption{Triangle anomaly coefficient as the area of a triangle on the toric diagram.}
\end{figure}

An explicit expression for the $a$-function was given in \cite{bz}.
First, a trial $R$-charge $h^I$ is assigned to each vertex of the toric diagram
subject to the constraint, $\sum_I h^I=2$.
The vertex $v_I$ is associated to a $D3$-brane wrapped on the calibrated
three-cycle $\S^I$ in $Y$ through $v_I \cdot y =0$.
Then, the $R$-charge of $\Phi_{IJ}$ is $R(\Phi_{IJ}) = \sum_{K=I+1}^J h^K$
or $R(\Phi_{IJ}) = \sum_{K=J+1}^{I} h^K$ depending on the sign of
$\bra w_I, w_J \ket$. The trial $a$-function is given by \cite{bz}
\be
\frac{32}{9} a = C_{IJK} h^I h^J h^K
&=& F \left( \thalf \sum h^I \right)^3 +
\sum_{I<J} \bra w_I, w_J \ket
\left( \sum_{K=I+1}^J h^K - \thalf \sum h^I \right)^3
\nn \\
&\equiv& F x^3 + \sum_{I<J} \bra w_I, w_J \ket
\left( y_{IJ}- x \right)^3 .
\ee
The first term is the contribution of gaugini while
the other terms account for the fermionic components of $\Phi_{IJ}^r$.
We replaced 1's appearing in the formula of \cite{bz} by $\thalf \sum_I h^I$
using the constraint $\sum_I h^I=2$
as we want to express $a$ as a homogeneous cubic function of $h^I$'s and
read off the anomaly coefficients.

In the simplest case, $d=3$, we can check (\ref{ccc}) explicitly,
\be
\frac{32}{9}a &=& F (x^3 + (h^1-x)^3 + (h^2 - x)^3 + (h^3 - x)^3 )
\nn \\
&=& 3 F h^1 h^2 h^3 = 6 \times \half |\bra v_1, v_2, v_3 \ket| h^1 h^2 h^3 ,
\ee
where we used $\bra w_1 ,w_2 \ket = \bra w_2 ,w_3 \ket= \bra w_3 ,w_1 \ket =
|\bra v_1, v_2, v_3 \ket| = F$. Now, we proceed by induction.
Assume the relation (\ref{ccc}) holds for a toric diagram
with $d$ vertices, and then add another vertex $v_{d+1}$. We distinguish
the objects for the new diagram by putting tilde above them.
\be
\frac{32}{9}\tilde{a} &=& \tilde{F} \tilde{x}^3 +
\sum_{I<J}^{d+1} \bra \tilde{w}_I, \tilde{w}_J \ket
\left( \tilde{y}_{IJ}- \tilde{x} \right)^3
\nn \\
&=&
(F+\bra \tilde{w}_d, \tilde{w}_{d+1} \ket ) (x + \thalf h^{d+1})^3
+ \sum_{I<J}^{d-1} \bra w_I, w_J \ket (y_{IJ} - x - \thalf h^{d+1})^3
\\
&&+ \sum_{I=1}^{d-1} \bra w_I, \tw_{d} \ket (y_{Id} - x - \thalf
h^{d+1})^3 + \sum_{I=1}^{d-1} \bra w_I, \tw_{d+1} \ket (y_{Id} - x +
\thalf h^{d+1})^3 \
\nn \\
&&+ \bra \tilde{w}_d, \tilde{w}_{d+1} \ket (-x + \thalf h^{d+1})^3 .
\nn
\ee
By collecting terms with $(h^{d+1})^n$ ($n=0,1,2,3$),
one can show that (\ref{ccc}) holds for all $d+1$ vertices.
\footnote{
We thank Eunkyung Koh for carrying out this `forward' proof completely.
}
The simplest one turns out to be the $(h^{d+1})^0$ term.
Setting $h^{d+1}=0$, we readily find
\be
\label{induc}
\tilde{a}|_{h^{d+1}=0} = a,
\ee
since the $\bra \tilde{w}_d, \tilde{w}_{d+1} \ket$ terms cancel out
and  $\bra w_I, \tw_{d} \ket + \bra w_I, \tw_{d+1} \ket = \bra w_I,
w_{d} \ket$. In fact, we can use (\ref{induc}) to reverse the
direction of the mathematical induction. That is, we can begin with
$d>3$ vertices and choose any three for which we want to compute
$C_{IJK}$. Then (\ref{induc}) allows us to remove the rest of
the vertices successively until we finally reach $d=3$. The value of
$C_{IJK}$ does not depend on the other vertices.

\subsection{Applications}

To demonstrate the utility of the compact formula (\ref{ccc}), we shall
now apply it to rederive two known results.

First, let us show that the triangle `t Hooft anomaly of baryon
symmetries always vanishes \cite{bz}: $\Tr B^3 = C_{IJK} B^I B^J B^K
= 0$, where $B^I$ is an arbitrary linear combination of the baryon
charges only: $B^I = t^a Q_a^I$. For example, when $d=4$,
\be
\frac{1}{3} C_{IJK}B^I B^J B^K &=& \bra B^1 v_1, B^2 v_2, B^3 v_3
\ket + \bra 2 , 3, 4\ket + \bra 3, 4, 1 \ket + \bra 4, 1, 2 \ket
\nn
\\
&=& \bra (1 + 2 + 3 + 4 ), 2, 3 \ket + \bra (1 + 2 + 3 + 4 ), 4, 1 \ket
\nn
\\
&=& 0.
\ee
In the last step, we used the toric relation $\sum_I Q_a^I v_I^i = 0$.
Similarly, for arbitrary $d$, vanishing of $\Tr B^3$ follows from
$\sum_I B^I ( B^J B^K \bra v_I, v_J, v_K \ket)  = 0$ (no sum over  $J,
K$). We relegate the general proof to appendix A.

\begin{figure}[htbp]
   \centering
   \includegraphics[width=8cm]{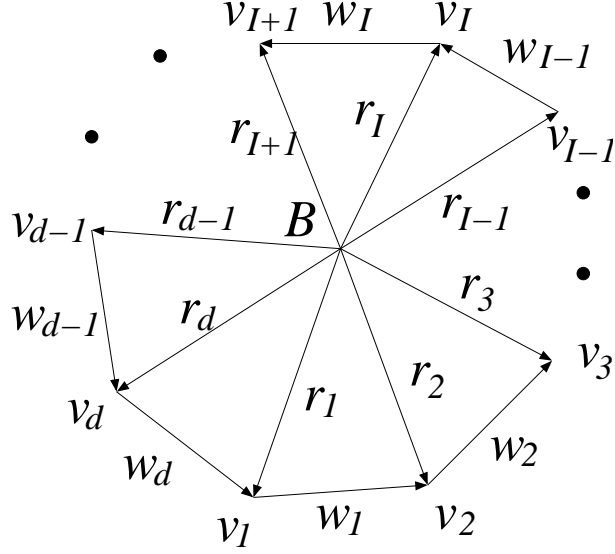}
   \caption{The Reeb vector as a point $B$ inside the polygon \cite{bz}.}
\end{figure}

Second, let us show the equivalence of $a$-maximization in
a quiver gauge theory and $Z$-minimization of the dual toric SE manifold
proposed in \cite{msy} and proven in \cite{bz}. Following \cite{bz}, we
parameterize the Reeb vector by $(b^1, b^2, b^3) = 3(1,x^2 ,x^3)$ and
define
\be r_I = (x^2 ,x^3) - (v_I^2, v_I^3), && A_I = \bra r_I, w_I
\ket,
\\
L^I(x^2,x^3) = \frac{\bra w_{I-1}, w_I\ket}{A_{I-1}A_I},&& S = \sum_I
L^I .
\ee
Then the results of \cite{msy} can be translated to the following
forms of trial $R$-charges and $a$-function:
\be
h^I_{MSY} \equiv \frac{2 L^I}{S} \qquad \mbox{and} \qquad
a_{MSY} = \frac{9}{32} \left( \frac{24}{S}\right) .
\ee
In \cite{bz}, it was shown that maximization of $a_{CFT}$ with
respect to trial $R$ charges is equivalent to maximization of
$a_{MSY}$ with respect to the Reeb vector components $(x^2,x^3)$.
The first step of the proof asserts that the baryon charges decouple
from the maximization process:
\be \Tr B R^2 |_{h^I = h^I_{MSY}} = 0 \;\;\;\;\;  \iff \;\;\;\;\;
C_{IJK} B^I L^J L^K = 0. \label{bll} \ee
Then it remains to prove that the maximization process yields the
same result. In fact, $a_{CFT}$ and $a_{MSY}$ are shown to be equal
even before maximization:
\be a_{CFT}|_{h^I = h^I_{MSY}} = a_{MSY} \;\;\;\;\; \iff \;\;\;\;
C_{IJK} L^I L^J L^K = 3 S^2. \label{lll} \ee
A complete proof of these two steps were presented in the (rather
long) appendix of \cite{bz}.

Here we note that (\ref{ccc}) offers a shorter and perhaps more
intuitive proof. As we prove in the appendix, both of the above
statements follow from a single lemma:
\be c_I \equiv C_{IJK} L^J L^K = 3 S + \bra r_I , u \ket ,
\label{lemma}\ee
where $u$ is some vector independent of the label $I$. If the lemma
is true, (\ref{bll}) follows from $\sum_I Q_a^I = 0 = \sum_I Q_a^I
v_I$ and (\ref{lll}) from $\sum_I L^I r_I = 0$. The proof of the
lemma is quite straightforward if we combine (\ref{ccc}) with the
original reasoning of \cite{bz}. See appendix B.

\subsection{More on the flavor charges and decoupling}

In gauge theory, we maximize the $a$-function
\be
a = \frac{9}{32} C_{IJK} h^I h^J h^K ,
\ee
subject to the constraint $\sum h^I = 2$. As the $R$-charge is a
linear combination of baryon and flavor charges, we can write
\be
\label{varch} h^I = t^a Q_a^I  + s^i F_i^I .
\ee
In this new basis, the constraint means $s^1 =2$,
as one can see from the extended toric relation (\ref{extor})
and $v_I^1=1$.
In fact, $s^i$ are related to the Reeb vector simply as $s^i =
(2/3)b^i$. At this stage, as discussed in section 2.1, $F_i^I$ is
ambiguous. The values of $t^a$ at the maximum of the $a$-function
depend on the choice of $F_i^I$, while the values of $s^i$ and
the $a$-function do not.

As discussed less explicitly in \cite{bz}, we can perform the
maximization process in two steps.
\be
\bar{a}(s,t) \equiv \frac{1}{3} C_{IJK} h^I h^J h^K
&=&
C_{i ab} s^i t^a t^b +  C_{ij a} s^i s^j t^a + \frac{1}{3} C_{ijk} s^i s^j s^k
\\
&\equiv&
m_{ab}(s) t^a t^b + 2n_a(s) t^a + \frac{1}{3} C_{ijk} s^i s^j s^k .
\ee
This is a quadratic function of $t^a$, so
maximization with respect to $t^a$ is done trivially to give
$\bar{t}^a(s) = -m^{ab}(s) n_b(s)$. Inserting it back to (\ref{varch}),
\be
\label{hbar}
\bar{h}^I(s) &=& -Q_a^I m^{ab}(s) n_b(s) + F_i^I s^i ,
\\
\bar{a}(s) &=& - m^{ab}(s) n_a(s)n_b(s)+ \frac{1}{3} C_{ijk} s^i s^j s^k .
\ee
The result discussed in the last subsection suggests the following
identification: \be \label{newpf} h^I_{MSY}(x^2, x^3) =
\bar{h}^I(s)|_{s=2(1,x^2,x^3)}, \;\;\;\;\; a_{MSY}(x^2,x^3) =
\frac{27}{32} \bar{a}(s)|_{s=2(1,x^2,x^3)}. \ee We checked
explicitly that this relation holds in many examples. If proven in
general, (\ref{newpf}) will establish the equivalence $a_{CFT} =
a_{MSY}$ in a somewhat more direct way than the approach of
\cite{bz} rederived in section 2.3.

For the rest of this paper, we shall assume that (\ref{newpf})
holds, and examine its implications. It is convenient to reinstate
the $s^1$-dependence of the quantities we defined earlier. For
example,
\be \bar{L}^I(s) \equiv \frac{\langle v_{I-1},v_I,v_{I+1} \rangle
}{\langle s,v_{I-1},v_I \rangle \langle s, v_I, v_{I+1}\rangle},
\;\;\;\;\; \bar{S}(s) \equiv \frac{1}{s^1} \sum_I \bar{L}^I(s),
\;\;\;\;\; h_{MSY}^I(s) \equiv \frac{\bar{L}^I(s)}{\bar{S}(s)}. \ee
Note that $h^I(s)$ satisfies
\be h^I(s) v_I^j = s^j . \ee
For $\bar{h}^I$, this holds due to the toric relation (\ref{extor}),
while for $h_{MSY}^I$ it has a geometric explanation, which we
review in appendix \ref{appB}. Differentiating, we find \be
\frac{\partial h^I}{\partial s^i} v_I^j = \d_i^j. \ee Thus
$(\partial h^I/\partial s^i)$ satisfy the same relation as $F_i^I$
in (\ref{extor}). We therefore define the `canonical' flavor charge
as
\be \label{fcan} \hat{F}_i^I \equiv \left. \frac{\partial
h^I}{\partial s^i}\right|_{s=s_*} \qquad (i=1,2,3), \ee
where $s_*$ denotes the value of $s$
which maximizes the $a$-function.
We will show in the next
section that this is precisely the flavor charge captured by
supergravity.

An important feature of the canonical flavor charge is that it makes
$\hat{C}_{Ri a} \equiv s_*^j \hat{F}_j^I \hat{F}_i^J Q_a^K C_{IJK}$
vanish. Suppose we work in the `canonical frame', that is, we
substitute $\hat{F}_i^I$ for $F_i^I$ in (\ref{varch}) and proceed.
Since $h^I$ is a homogeneous function of $s$ of degree 1, we can
always write $h^I(s) = s^i \frac{\partial h^I}{\partial s^i}$. In
the canonical frame, this implies that $\bar{t}^a(s_*) = 0$. Next,
by differentiating (\ref{hbar}) in the canonical frame, and
recalling (\ref{fcan}),
\be
\frac{\partial h^I}{\partial s^i} = Q_a^I \frac{\partial
\bar{t}^a}{\partial s^i} + \hat{F}_i^I
\qquad \imp \qquad
\left. \frac{\partial \bar{t}^a}{\partial s}\right|_{s_*} = 0.
\ee
Now, combining $\bar{t}^a (s_*) = 0 = \frac{\partial
\bar{t}^a}{\partial s^i}|_{s_*}$ with $\bar{t}^a(s) = -m^{ab}(s)
n_b(s)$, we find that
\be \label{rai1} \left. \frac{\partial n_a}{\partial s^i}
\right|_{s_*} = C_{aij } \, s_*^j = {C}_{R ai} = 0. \ee
This demonstrates the decoupling property among the global charges.


\section{Comparison with Supergravity}

In this section we compare our main results from the previous section
with the supergravity computation. First, we work out the KK reduction
at the linearized level. It was already done in \cite{bgiw} where
a covariant action in ten dimensions was assumed.
To avoid the usual difficulty with the self-dual five form of
IIB supergravity,
we follow the common path \cite{krvn, lmsr} of using only the
equations of motion.

Second, we compare the flavor charges between field theory and
supergravity.
The agreement is perfect. We emphasize that both field theory
and supergravity
pick out a unique value of flavor charge and the mixing ambiguity is
resolved.

Finally, we would like to compare $C_{IJK}$ of field theory (\ref{ccc})
with supergravity by extending the KK reduction to the cubic order.
This has been carried out in a very recent paper \cite{bpt}.
In the last subsection of this paper, we check the relation
$\t_{IJ} = -3 C_{RIJ}$ \cite{trr} using the results of \cite{bgiw, bpt}
and find complete agreement.

\subsection{Massless vectors from linearized equations}

We shall follow the conventions of \cite{lmsr}. The IIB supergravity
equations of motion relevant to our analysis are
\be \label{fieldeq}
R_{m n} = \frac{4}{4!} F_{m i_1i_2i_3i_4} F_n{}^{i_1i_2i_3i_4},
\qquad F = * F, \qquad dF=0. \ee
In units in which the `radius' $l =
(4\pi^4 g_s N / {\rm Vol}(Y))^{1/4}l_s$ is set to be unity, the
background solution with $N$ units of $F$-flux is
\be
\label{background} ds^2 = ds^2_{AdS} + ds^2_{Y} \qquad \mbox{and}
\qquad F = \vol_{AdS} + \vol_{Y}. \ee
The metric is normalized such that $R_{\m\n} = -4g_{\m\n}$ for
$AdS_5$ and $R_{\a\b} = +4g_{\a\b}$ for $Y$. We shall now perturb
around the background solution and obtain equations of motion for
massless vector gauge fields up to linear order.

The gauge fields for baryon symmetries arise from fluctuations of
the RR five-form field strength,
\be
\d F = F^a \wedge \w_a - *F^a \wedge *\w_a ,
\ee
around the background (\ref{background}). The second term ensures
that the self-duality constraint $F=*F$ is satisfied. Here, the
Hodge duals are factorized to AdS$_5$ and $Y$, respectively. At the
linearized level, no other perturbation is needed.

The gauge fields for flavor symmetries arise from fluctuations
along the isometries. We take the following ansatz for the fluctuations:
\be
ds^2 &=& ds^2_{AdS}
+ g_{\a\b}(dy^\a + K_i^\a A^i) (dy^\b + K_j^\b A^j),
\\
F &=& \mbox{vol}_{AdS} + \mbox{vol}_{Y}+ d C, \;\;\;\;\; C =
\frac{1}{8} (B^i \wedge *dK_i + *dB^i \wedge K_i) . \ee
The metric part of the ansatz is the standard one in KK reduction.
The vector $B^i$ from the RR five-form field-strength must be turned
on also because $A^i$ and $B^i$ mix already at linearized order
\cite{krvn}. As the ansatz for $F$ is written in terms of the
potential $C$, the Bianchi identity holds automatically.  Again, the
Hodge duals are factorized to AdS$_5$ and $Y$, respectively.

The mixed components of the Einstein equation and the self-duality
equation give, respectively,
\be
(\square - 8) A^i = (\square +8 ) B^i \qquad \mbox{and} \qquad
(\square - 8) B^i = 8 A^i,
\ee
where we defined $\square \equiv (*d*d)_{AdS}$. We also used the fact that
$d*K_i=0$, $d*dK_i = 8 *K_i$ on $Y$, which follows from the Killing equation
$\nabla_\a K_\b + \nabla_\b K_\a =0$ and $R_{\a\b}= 4 g_{\a\b}$.
We can easily diagonalize the two equations to obtain the mass
eigenstates:
\be
\square (A^i + B^i) = 24 (A^i + B^i ),
\;\;\;\;\;
\square (A^i - 2B^i) = 0.
\ee
To keep the massless fields only, we set $B^i = -A^i$.

Now, we can read off the gauge kinetic term of the massless gauge
fields from the AdS$_5$ components  of the field equations
(\ref{fieldeq}). They yield via AdS/CFT the coefficients $\t_{IJ}$
of the two-point correlators for conserved global currents $J_I$ in
gauge theory. The result is to be compared with \cite{bgiw}. A
precise comparison, however, requires normalization of the gauge
fields, which is related to the normalization of the charges on the
gauge theory side. So, we shall first discuss how to find the
correct normalization.

\subsection{Charges}

As stated in (\ref{baryoncharge}),
a natural normalization for the baryon charges is
\be
\label{bch}
B_a\left[\S^I\right] = \int_{\S^I} \w_a = Q_a^I ,
\ee
where $\{ \w_a\}$ form an integral basis of $H^3(Y,\IR)$.
The KK analysis of the previous subsection suggests that
the flavor charges can be measured with the replacement of
$\w_a$ by $*dK_i$ modulo an arbitrary multiplicative constants.
The correct normalization turns out to be
\be
\label{fch}
F_i^I = \frac{2\pi}{V} \int_{\S^I} (*dK_i)/8 \qquad \qquad  (i=1,2,3),
\ee
where $V$ denotes $\Vol(Y)$. As a first check, note that the
R-charge is given by
\be
\label{qcheck}
R^I = \frac{2}{3} b^i F_i^I = \frac{\pi}{6 V} \int_{\S^I} *d K_R
= \frac{\pi}{3 V} \Vol(\S^I),
\ee
in agreement with the well-known result in the literature
\cite{bhk}. Note that we are abusing the notations a bit and use
$K_i$ to denote both the Killing vector and its dual one-form. In
the last step of (\ref{qcheck}), we used the local $U(1)_R$
fibration description of the SE manifold $Y$ (see also
(\ref{relations})):
\be
\label{yb1}
ds_Y^2 = (e^0)^2 + ds_B^2,
&&
e^0 \equiv \frac{1}{3} d\psi + \s,
\;\;\;\;\;
K_R = 3 \frac{\partial}{\partial \psi},
\\
\label{yb2}
R_{\m\n}^{(B)} = 6 g_{\m\n}^{(B)},\qquad
&&
d\s = 2 J_B,
\; \qquad
\vol_\S = e^0 \wedge J_B.
\ee
It is instructive to compare (\ref{fch}) with known results.
On the supergravity side, generalizing the
analysis for the $R$-charge in \cite{bhk}, the authors of \cite{bgiw}
showed that, for {\em non-}$R$ flavor charges,
\be
\label{fbgiw}
F_i^I = - \frac{\pi}{V} \int_{\S^I} (i_{K_i} \s) \vol_{\S}
= - \frac{2\pi}{V} \int_{\S^I} y_i \; \vol_\S
= - \frac{1}{V} \frac{\partial V}{\partial v_I^i} \qquad (i=2,3),
\ee
where in the last expression, the volume $V$ is regarded as
a function of
the toric data $v_I^i$. On the other hand, as we reviewed in the
last section
the field theory result is
\be
\label{fbz}
F_i^I = \half \frac{\partial}{\partial x^i} h^I_{MSY} (\vec{x})
\qquad (i=2,3).
\ee

We now show that all three expressions for the flavor charges
(\ref{fch}), (\ref{fbgiw}) and (\ref{fbz}) are in fact the same.
To see (\ref{fbz}) is the same as the last expression in (\ref{fbgiw}),
we note that
\be
\label{aaa}
h_{MSY}^I = \frac{2L^I}{S},
\qquad
\left. \frac{\partial V}{\partial x^i}\right|_{x_*}  = 0 ,
\qquad
\frac{\partial S}{\partial v_I^i} = - \frac{\partial L^I}{\partial x^i},
\ee
where $x_*$ denotes the value of $\vec{x}$ that minimizes
$S$ which is proportional to $V=\Vol(Y)$.
The last identity in (\ref{aaa}) holds for arbitrary values of $\vec{x}$,
as can be checked by explicit computation.

To see that the first expression in (\ref{fbgiw}) is the same as (\ref{fch}),
it suffices to show the equality:
\be
\int_{\S^I} *_5 dK_i = -4 \int_{\S^I} (i_{K_i}\s) \vol_\S .
\ee
This can be proven using (\ref{yb1}), (\ref{yb2}). The one-form dual
to the flavor Killing vector $K_i = \partial/\partial \phi^i$
($i=2,3$) can be decomposed into the base $B$ and the local $U(1)_R$
fiber:
\be
K_i = \bar{K}_i + (i_{K_i} \s) e^0
\qquad \mbox{such that} \qquad
dK_i = d \bar{K}_i + 2 (i_{K_i} \s) J_B - 2(i_{K_i}J_B) e^0 .
\ee
Here, the relation $\CL_{K_i} \s \equiv d(i_{K_i}\s) + i_{K_i}(d\s)
= 0$ was used. Splitting the three-cycle $\S^I$ into the $U(1)_R$
fiber and a 2-cycle $B^I$ in the base $B$,
\be
\int_{\S^I} *_5 dK_i = \int e^0 \int_{B^I}*_4d\bar{K}_i
+ 2 \int_{\S^I} (i_{K_i}\s) \vol_\S .
\ee
The final step of the proof follows from the identity:
\be
\label{siden}
d\bar{K}_i + *_4 d\bar{K}_i = -6(i_{K_i} \s) J_B .
\ee
The left-hand side of (\ref{siden}) is manifestly a self-dual
$(1,1)$ form, so it must be proportional to the K\"ahler form $J_B$.
To see if (\ref{siden}) is consistent, take an exterior derivative
to (\ref{siden}). We find that $d *_4 d\bar{K}_i = 12 *_4 \bar{K}_i$
from the left-hand side is indeed equal to
\[
-6 \,d(i_{K_i} \s) \wedge J_B = 12 (i_{K_i}J_B)\wedge J_B = 12 *_4 \bar{K}_i
\]
from the right-hand side. This still leaves a room for a term
proportional to the K\"ahler form $J_B$ on the right-hand side of
(\ref{siden}). To show that such a term does not appear, let us now
integrate (\ref{siden})  over the base $B$. The left-hand side
vanishes by integration parts  and $dJ=0$, while
\be
\int_{B} (i_{K_i} \s) \;\;\;\propto \;\;\; \int_{Y} (i_{K_i} \s)
\;\;\; \propto \;\;\;
\frac{\partial V}{\partial b^i} =0,
\ee
as a result of volume-minimization \cite{msy, bgiw}.

\subsection{Gauge kinetic coefficient $\tau_{IJ}$ revisited}

With the normalization for the flavor charges fixed,
from the KK reduction analysis in section 3.1, we can  compute the gauge field kinetic term coefficient $\t_{IJ}$ and compare them with \cite{bgiw}.
To do so in uniform manner along with the flavor charges (\ref{fch}), we rescale the harmonic three-forms
by $2\pi/V$ relative to (\ref{bch}), viz.
\be
\label{newnorm}
\frac{2\pi}{V} \int_{\S^I} \w_a = Q_a^I. \ee
Then, the expressions for $\t_{IJ}$ are
\be
\label{tow}
\t_{ab} = \frac{16 \pi^3}{V^2} \int_{Y} \w_a \wedge *\w_b, \qquad
\t_{ai } = 0, \qquad
\t_{ij} = \frac{3\pi^3}{V^2} \int_{Y} K_i \wedge *K_j.
\ee
The baryon components $\t_{ab}$ are precisely the same as in
\cite{bgiw}. As for the flavor components, the coefficient of
gravi-photon ($R$-symmetry) is
\be
\t_{RR} = \left(\frac{2}{3}\right)^2 b^i b^j \t_{ij}
=
\frac{3\pi^3}{V^2} \left(\frac{2}{3}\right)^2 \int_{Y_5} K_R \wedge * K_R
=\frac{4\pi^3}{3V} = \frac{16}{3} a ,
\ee
in agreement with \cite{bgiw}.
For the other flavor symmetries, the expression from \cite{bgiw}
looks slightly different:
\be
\label{tow2}
\t_{ij} = \frac{12\pi^3}{V^2} \int_{Y} (i_{K_i}\s)(i_{K_j}\s) \vol_{Y}
\qquad \quad
(i,j = 2,3).
\ee
It agrees with (\ref{tow}) if and only if
\be
\label{abc}
\int_{Y_5} K_i \wedge *K_j = 4 \int_{Y} (i_{K_i}\s)(i_{K_j}\s) \vol_{Y}
\qquad
(i,j = 2,3).
\ee
This identity was stated in \cite{bgiw} without proof.
We note that it can be
verified using (\ref{siden}),
and other relations we used in section 3.2.
See appendix C for details.

\subsection{Chern-Simons coupling $C_{IJK}$}

The Chern-Simons coupling $C_{IJK}$ is obtainable in KK reduction by
using the ansatz of subsection 3.1 and computing the fluctuation up
to cubic order along the line of \cite{lmsr, af, me}. While this
work was in progress, Ref. \cite{bpt} appeared, where the full
computation was performed using a slightly different approach. The
difference is that our ansatz manifestly satisfy $dF=0$ but the
self-duality equation is non-trivial, while an alternative ansatz
was used in \cite{bpt}, where $F$ is manifestly self-dual but not
necessarily closed.

A central step in \cite{bpt} was to combine the baryon symmetries
and flavor symmetries together into some three-forms $\w_I$ such
that
\be
\int_{\S^I} \w_J = \d_J^I.
\ee
Comparing with our charge normalizations (\ref{fch}),
(\ref{newnorm}) and
the toric relation (\ref{extor}), we find that
\be
\w_I = \frac{2\pi}{V} (u_I^a \w_a + v_I^i *dK_i/8).
\ee
We can use it to re-express the result of \cite{bpt} in
a more convenient form:
\be
\label{cscoeff}
C_{ijk} &=& \frac{3\pi^3}{8 V^2} \int_Y K_i \wedge d K_j \wedge dK_k ,
\nonumber \\
C_{ija} &=& \frac{2 \pi^3}{V^2} \int_Y *(K_i d K_j )  \wedge \w_a ,
\nonumber \\
C_{i ab} &=& \frac{8\pi^3}{V^2} \int_Y  \w_a \wedge  i_{K_i}\w_b.
\ee
As a consistency check, we compute the $a$-function, which is
proportional to $C_{RRR}$, and obtain the expected result:
\be
a = \frac{9}{32} C_{ijk} b^i b^j b^k \left(\frac{2}{3}\right)^3 &=&
\frac{\pi^3}{32 V^2} \int_{Y} K_R \wedge dK_R \wedge dK_R \nonumber
\\
&=& \frac{\pi^3}{32 V^2} \int_{Y}  e^0 \wedge (2J_B)\wedge (2J_B) =
\frac{\pi^3}{4 V}.
\ee
%

\subsection{$\t_{IJ} = - 3C_{RIJ}$ relations}
Utilizing  the supergravity expressions for the gauge kinetic
coefficients (\ref{tow}) and the Chern-Simons coefficients
(\ref{cscoeff}), we can now demonstrate the relation suggested in
\cite{trr} between the two-point correlators and the triangle `t
Hooft anomalies involving conserved currents in the gauge theory:
\be \t_{IJ} = -3 \Tr R\, F_I \,F_J \equiv -3 C_{RIJ}. \ee
Here, $F_I$ include both baryon and non-$R$ flavor charges.

First, $\t_{ab}=-3 C_{R ab}$ follows from the fact that,
in the local $U(1)_R$ fibration description of $Y$ given in
(\ref{yb1}), (\ref{yb2}),
$\w_a = e^0 \eta_a$ for some {\sl anti}-self-dual two-form
$\eta_a$ on $B$
\cite{bgiw}.
Next, $\t_{ia}=0$ implies that $C_{R ia}$ must also vanish.
It is indeed so
because $K_R = e^0$, $\w_a = e^0 \eta_a$ as mentioned above,
and $\w_a$
is harmonic. This also agrees with the field theory computation
(\ref{rai1}).
The last  relation $\t_{ij}= -3C_{R ij}$ amounts to
\be
\label{def}
\int K_R \wedge dK_i \wedge dK_j = -4 \int K_i \wedge *K_j.
\ee
This simply follows from (\ref{abc}), as explained in appendix C.

\vskip 1cm

\centerline{\bf Acknowledgement}

\vskip .5cm

We are grateful to David Berenstein, Nakwoo Kim, Seok Kim, Kimyeong
Lee, Oleg Lunin, Ho-Ung Yee, Piljin Yi and Sang-Heon Yi for useful
discussions, and Eunkyung Koh for collaboration on parts of section
2. SML was supported in part by the Faculty Grant of Seoul National
University. SJR was supported in part by the KRF Leading Scientist
Grant, the KRF Star Faculty Grant, and the KOSEF SRC Program through
``Center for Quantum Spacetime" (R11-2005-021).

\section*{Appendix}

\appendix

\section{$\Tr (B^3) = 0$}

We prove that $C_{IJK}B^I B^J B^K = 0$ for any linear combination of
baryon symmetries. The proof consists of a combination of our
formula $C_{IJK} = |\bra v_I, v_J, v_K \ket|/2$, the toric relation
$B^I v_I = 0$, and some combinatoric manipulations. More concretely,
we show that
\be 0 &=& \half \sum_{(J,K)} \left[ \sum_I \bra v_I,
v_J, v_K \ket B^I B^J B^K \times (d - 2 (K-J)) \right]
\nn \\
\label{qqq} &=& \sum_{(J,K)} \left[ \sum_I (-1)^{(I,J,K)} C_{IJK}
B^I B^J B^K \times (d - 2 (K-J)) \right]
\\
&=& \frac{d}{6} \sum_{I,J,K} C_{IJK} B^I B^J B^K. \nn \ee
The notations require some clarification. The $(J,K)$ sum runs over
all possible pairs with $0 < K-J \le d/2$ (mod $d$). The $I$ sum
then runs over all vertices. The first line is a trivial consequence
of $B^I v_I = 0$. The second line simply says that $C_{IJK}$ is
equal to $\bra v_I, v_J ,v_K \ket/2$ up to a sign depending on
whether $I$ lies on the long(+) or short(-) path between $J$ and
$K$. The weight factor $d-2(K-J)$ ensures that if we choose some
fixed triangle $(I,J,K)$ and collect all terms proportional to
$C_{IJK}$ from the second line, the net coefficient always turns out
to be $d$, independent of the choice of the triangle.
\begin{figure}[htbp]
   \centering
   \includegraphics[width=7cm]{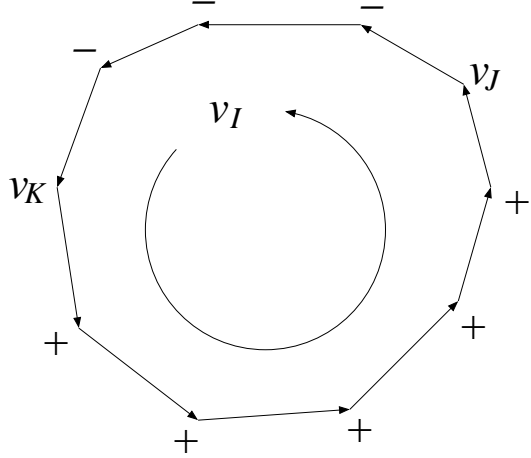}
   \caption{The sign assignment in the second line of (\ref{qqq}).}
\end{figure}

Let us check the last statement. Let $l_1$, $l_2$, $l_3$ be the
number of edges between $(I,J)$, $(J,K)$ and $(K,I)$ respectively,
so that  $l_1 +l_2 +l_3 = d$. Without loss
of generality, we may assume that $l_1 \le l_2 \le l_3$. We collect
the terms in two separate cases:
\benu

\item
$l_3 \le d/2$: The sign is positive for all three contributions from
the second line of (\ref{qqq}). The net coefficient is
$(d-2l_1)+(d-2l_2)+(d-2l_3) = d$.

\item
$l_3 > d/2$: The sign is negative in one of the three contributions
from the second line of (\ref{qqq}). The net coefficient is
$(d-2l_1)+(d-2l_2)-(d-2(d-l_3)) = d$.

\eenu
This completes the proof.

\section{Equality of $a_{CFT}$ and $a_{MSY}$ \label{appB}}

We prove the lemma (\ref{lemma}):
\be
c_1 \equiv C_{IJK} L^J L^K =
3S + \bra r_I, u \ket.
\ee
As explained in section 2.3, this lemma is sufficient to establish
the equality between $a_{CFT}$ and $a_{MSY}$. The main idea for the
proof is the same as in the original one \cite{bz}, but  our formula
$C_{IJK} = |\bra v_I, v_J, v_K \ket|/2$ simplifies the computation
involved considerably.

The definition of $w_I$, $r_I$, etc. are the same as in section 2.3.
In what follows, we will need the following identity \cite{bz}: \be
\label{biden} L^I r_I = \frac{w_{I-1}}{A_{I-1}} - \frac{w_I}{A_I},
\;\;\;\; \imp \;\;\;\; \sum_I L^I r_I = 0 \;\;\;\; \mbox{or}
\;\;\;\; \sum_I L^I v_I = (1,x^2,x^3)\sum_I L^I. \ee Geometrically,
the last equation follows from integrating the `gradient of a
constant function' over the polyhedral cone $\D$ and applying
Stokes' theorem; see (2.91) of \cite{msy}.

Getting back to the lemma, we write $c_1$ as
\be c_1 = \sum_{2 \goto d} \bra v_1, v_J, v_K \ket L^J L^K. \ee
Here the notation ($2\goto d$) means that the
sum is taken over $2 \le J < K \le d$. In the following, we will use
notations like ($2\goto 1$), which means the range $2 \le J < K \le
d+1$ with $v_{d+1} \equiv v_{1}$.

As in \cite{bz}, we first compute the difference between two
adjacent $c_I$'s. Using the relation $ \bra v_I, v_J, v_K \ket =
\bra r_I, r_J \ket + \bra r_J, r_K \ket + \bra r_K, r_I \ket, $ we
find, for example,
\be c_2 - c_1 = \bra w_1, u_1 \ket, \;\;\;\;\;
u_1 \equiv  \sum_{2 \goto 1} ( r_J -   r_K) L^J L^K -2S
\frac{w_1}{A_1}. \ee
The second term in the definition of $u_1$ does
not affect the value of $c_2 -c_1$. We include it (and similar terms
for all $u_I$) to make all the $u_I$'s the same ($u_1 = u_2 = \cdots
= u_d \equiv u$) :
\be u_2 - u_1 &=& - 2 \sum_{3\goto 1} (r_2 - r_K)
L^2 L^K - 2S \left( \frac{w_2}{A_2} - \frac{w_1}{A_1} \right)
\nn \\
&=& -2 \left[ r_2 L^2 (S - L^2) + r_2 (L^2)^2 \right] - 2 S
\left(\frac{w_2}{A_2} - \frac{w_1}{A_1} \right) = 0 , \ee
where we used (\ref{biden}). This implies that $c_I - \bra r_I, u
\ket$ is independent of the index $I$. Performing the subtraction
and using (\ref{biden}) once again, we find
\be c_1 - \bra r_1, u
\ket = 2S + \sum_{2\goto d} \bra r_J, r_K \ket L^J L^K \equiv 2S +T
. \ee
Finally, we show that $T=S$ by mathematical induction. To
begin with, we note that for $d=3$,
\be T =  \bra r_2, r_3 \ket L^2
L^3 = A_2 \times \frac{\bra w_1 ,w_2 \ket}{A_1 A_2} \times
\frac{\bra w_2 ,w_3 \ket}{A_2 A_3}
= \frac{\bra w_1 ,w_2 \ket}{A_1 A_2} + \frac{\bra w_2 ,w_3 \ket}{A_2
A_3} + \frac{\bra w_3 ,w_1 \ket}{A_3 A_1}  = S, \ee
where we used the fact that, when $d=3$, $\bra w_1 ,w_2 \ket = \bra
w_2 ,w_3 \ket= \bra w_3 ,w_1 \ket = A_1+ A_2 + A_3$. Now, assume
that $T=S$ holds for a toric diagram with $d$ vertices. As we add
another vertex $v_{d+1}$, most of the terms in $S$ and $T$ remain
unchanged. The only differences are
\be \tilde{S} - S &=&
\tilde{L}_{d}+ \tilde{L}_{d+1} + \tilde{L}_1 - (L_d + L_1) ,
\\
\tilde{T} - T &=& \bra r_d , r_1 \ket \tilde{L}^d \tilde{L}^1 + \bra
r_{d}, r_{d+1} \ket \tilde{L}^d \tilde{L}^{d+1} + \bra r_{d+1},
r_{1}  \ket \tilde{L}^{d+1} \tilde{L}^{1} - \bra r_{d}, r_{1} \ket
L^d L^1 , \ee
where we distinguished the objects for the new diagram
by adding tilde above them. Using the identity again (\ref{biden}),
we obtain
\be \bra r_{d}, r_{1}
\ket L^d L^1 &=& L_d + L_1 - \frac{\bra w_{d-1}, w_1 \ket}{A_{d-1}
A_1},
\\
\bra r_{d}, r_{d+1} \ket \tilde{L}^d \tilde{L}^{d+1}  &=&
\tilde{L}_{d}+ \tilde{L}_{d+1} - \frac{\bra w_{d-1}, \tw_{d+1}
\ket}{A_{d-1} \tilde{A}_{d+1}},
\\
\bra r_{d+1}, r_{1}  \ket \tilde{L}^{d+1} \tilde{L}^{1}  &=&
\tilde{L}_{d+1}+ \tilde{L}_{1} - \frac{\bra \tw_{d}, w_1
\ket}{\tilde{A}_{d} A_1},
\\
\bra r_d , r_1 \ket \tilde{L}^d \tilde{L}^1 &=& -\tilde{L}_{d+1} +
\frac{\bra w_{d-1}, \tw_{d+1} \ket}{A_{d-1} \tilde{A}_{d+1}} +
\frac{\bra \tw_{d}, w_1 \ket}{\tilde{A}_{d} A_1} - \frac{\bra
w_{d-1}, w_1 \ket}{A_{d-1} A_1}. \ee
Therefore, $T=S$ implies $\tilde{T}=\tilde{S}$. This completes the
proof.

\section{Some identities}

In this appendix, we prove two identities that we needed in section
3 to establish the relation between $\t_{ij}$ and $C_{ijk}$. Recall
that the one-form dual to the Killing vector $K_i$ is decomposed
under the local $U(1)_R$ fibration description of $Y$ (\ref{yb1}),
(\ref{yb2}) as
\be K_i = \bar{K}_i +
(i_{K_i} \s) e^0. \ee
The integral appearing in $\t_{ij}$ splits accordingly:
\be \frac{1}{2\pi} \int_{Y} K_i \wedge *_5 K_ j = \int_B \bar{K}_i
\wedge *_4 \bar{K}_j + \int_{B} (i_{K_i} \s) (i_{K_j} \s) \vol_{B}
\equiv A_{ij} + B_{ij}. \ee
The first identity (\ref{abc})
follows from a straightforward computation:
\be A_{ij} &=& \int_{B}
(i_{K_i}J_B) \wedge *_4 (i_{K_j}J_B) = -\thalf \int_{B} d(i_{K_i}\s)
\wedge *_4 (i_{K_j}J_B)
\nn \\
&=& \thalf \int_{B} (i_{K_i}\s) d*_4 (i_{K_j}J_B) = -\thalf\int_{B} (i_{K_i}\s)
d(\bar{K_j}\wedge J_B)
\\
&=& -\thalf \int_{B} (i_{K_i}\s)\left[ \thalf(d\bar{K}_j + *_4
d\bar{K}_j)+ \thalf(d\bar{K}_j - *_4 d\bar{K}_j) \right] \wedge J_B
\nn \\
&=&3 \int_{B} (i_{K_i}\s)(i_{K_j}\s)\thalf J_B\wedge J_B
\;\; = \;\;
3B_{ij} .
\nn
\ee
We used (\ref{siden}) in going from the third to the last line. The
second identity (\ref{def}) follows, since
\be \frac{1}{2\pi} \int_{Y} K_R \wedge dK_i \wedge dK_j &=& \int_{B}
(d\bar{K}_i + 2 (i_{K_i}\s)J_B) \wedge (d\bar{K}_j+ 2(i_{K_j}\s)J_B)
\\
&=& -8 A_{ij} + 8 B_{ij} = - 16 B_{ij} = -4  \left[ \frac{1}{2\pi}
\int_{Y_5} K_i \wedge *_5 K_ j \right].
\nn
\ee

\newpage



\begin{thebibliography}{99}



\bibitem{adscft}
J. M. Maldacena, ``The large $N$ limit of superconformal field
theories and supergravity," Adv. Theor. Math. Phys. {\bf 2} (1998)
231 [Int. J. Theor. Phys. {\bf 38} (1999) 1113]
[arXiv:hep-th/9711200].



\bibitem{kw}
I. R. Klebanov and E. Witten, ``Superconformal field theory on
threebranes at a Calabi-Yau singularity,Ó Nucl. Phys. B {\bf 536}
(1998) 199 [arXiv:hep-th/9807080].

\bibitem{gub}
S. S. Gubser, ``Einstein manifolds and conformal field theories,"
Phys. Rev. D {\bf 59} (1999) 025006 [arXiv:hep-th/9807164].

\bibitem{ach}
  B.~S.~Acharya, J.~M.~Figueroa-O'Farrill, C.~M.~Hull and B.~J.~Spence,
  ``Branes at conical singularities and holography,''
  Adv.\ Theor.\ Math.\ Phys.\  {\bf 2}, 1249 (1999)
  [arXiv:hep-th/9808014].


\bibitem{mp}
D.~R.~Morrison and M.~R.~Plesser,
  ``Non-spherical horizons. I,''
  Adv.\ Theor.\ Math.\ Phys.\  {\bf 3}, 1 (1999)
  [arXiv:hep-th/9810201].





\bibitem{gmsw1}
J. P. Gauntlett, D. Martelli, J. Sparks and D. Waldram,
``Sasaki-Einstein metrics on $S^2 \times S^3$ ," Adv. Theor. Math.
Phys. {\bf 8}, 711 (2004) [arXiv:hep-th/0403002].

\bibitem{clpp}
M. Cvetic, H. Lu, D. N. Page and C. N. Pope, ``New Einstein-Sasaki
spaces in five and higher dimensions," Phys. Rev. Lett. {\bf 95},
071101 (2005) [arXiv:hep-th/0504225].



\bibitem{ms1}
D.~Martelli and J.~Sparks,
  ``Toric geometry, Sasaki-Einstein manifolds and a new infinite class of
  AdS/CFT duals,''
  Commun.\ Math.\ Phys.\  {\bf 262}, 51 (2006)
  [arXiv:hep-th/0411238].

\bibitem{msy}
D. Martelli, J. Sparks and S. T. Yau, ``The geometric dual of
a-maximisation for toric Sasaki-Einstein manifolds,"
[arXiv:hep-th/0503183].

\bibitem{ms2}
D. Martelli and J. Sparks, ``Toric Sasaki-Einstein metrics on
$S^2\times S^3$," Phys. Lett. B {\bf 621}, 208 (2005)
[arXiv:hep-th/0505027].




\bibitem{bfhms}
S. Benvenuti, S. Franco, A. Hanany, D. Martelli and J. Sparks, ``An
infinite family of superconformal quiver gauge theories with
Sasaki-Einstein duals," JHEP {\bf 0506}, 064 (2005)
[arXiv:hep-th/0411264].

\bibitem{bk}
S. Benvenuti and M. Kruczenski, ``From Sasaki-Einstein spaces to
quivers via BPS geodesics: $L^{p,q|r}$ ," [arXiv:hep-th/0505206].

\bibitem{fhm}
S. Franco, A. Hanany, D. Martelli, J. Sparks, D. Vegh and B. Wecht,
 ``Gauge theories from toric geometry and brane tilings,"
[arXiv:hep-th/0505211].

\bibitem{bfz}
A. Butti, D. Forcella and A. Zaffaroni, ``The dual superconformal
theory for $L^{p,q,r}$ manifolds," JHEP {\bf 0509}, 018 (2005)
[arXiv:hep-th/0505220].






\bibitem{fhh}
B. Feng, A. Hanany and Y. H. He, ``D-brane gauge theories from toric
singularities and toric duality," Nucl. Phys. B {\bf 595}, 165
(2001) [arXiv:hep-th/0003085].

\bibitem{hiq}
A. Hanany and A. Iqbal, ``Quiver theories from D6-branes via mirror
symmetry," JHEP {\bf 0204}, 009 (2002) [arXiv:hep-th/0108137].

\bibitem{hw}
C. P. Herzog and J. Walcher, ``Dibaryons from exceptional
collections," JHEP {\bf 0309}, 060 (2003) [arXiv:hep-th/0306298].


\bibitem{fhkz}
S. Franco, A. Hanany and P. Kazakopoulos, ``Hidden exceptional
global symmetries  in 4d CFTs," JHEP {\bf 0407}, 060 (2004)
[arXiv:hep-th/0404065].


\bibitem{bhan}
S. Benvenuti and A. Hanany, ``New results on superconformal quivers,Ó
[arXiv:hep- th/0411262].

\bibitem{hken}
A. Hanany and K. D. Kennaway, ``Dimer models and toric diagrams,"
[arXiv:hep- th/0503149].

\bibitem{fhkvw}
S. Franco, A. Hanany, K. D. Kennaway, D. Vegh and B. Wecht,
 ``Brane
dimers and quiver gauge theories," [arXiv:hep-th/0504110].


\bibitem{bkr}
S. Benvenuti and M. Kruczenski, ``Semiclassical strings in
Sasaki-Einstein manifolds and long operators in $N = 1$ gauge
theories," [arXiv:hep-th/0505046].


\bibitem{hv}
A. Hanany and D. Vegh, ``Quivers, tilings, branes and rhombi,"
[arXiv:hep-th/0511063].

\bibitem{fhkv}
B. Feng, Y. H. He, K. D. Kennaway and C. Vafa, ``Dimer models from
mirror symmetry and quivering amoebae," [arXiv:hep-th/0511287].



\bibitem{cep}
 F. Canoura, J. D. Edelstein, L. A. Pando Zayas, A. V. Ramallo and D. Vaman,
``Supersymmetric branes on AdS$5 \times Y_{p,q}$ and their field theory
duals," [arXiv:hep-th/0512087].




\bibitem{amax}
K. Intriligator and B. Wecht, ``The exact superconformal
$R$-symmetry maximizes $a$," Nucl. Phys. B {\bf 667}, 183 (2003)
[arXiv:hep-th/0304128].


\bibitem{bbc}
M. Bertolini, F. Bigazzi and A. L. Cotrone, ``New checks and
subtleties for AdS/CFT and $a$-maximization," JHEP {\bf 0412}, 024
(2004) [arXiv:hep-th/0411249].


\bibitem{bz}
A. Butti and A. Zaffaroni, ``R-charges from toric diagrams and the
equivalence of $a$-maximization and $Z$-minimization," JHEP {\bf
0511}, 019 (2005) [arXiv:hep-th/0506232].

\bibitem{tachi}
 Y. Tachikawa,
 ``Five-dimensional supergravity dual of $a$-maximization,"
 [arXiv:hep-th/0507057].

\bibitem{bgiw}
E. Barnes, E. Gorbatov, K. Intriligator and J. Wright, ``Current
correlators and AdS/CFT geometry," [arXiv:hep-th/0507146].

\bibitem{bz2}
A. Butti and A. Zaffaroni, ``From toric geometry to quiver gauge
theory: The equivalence of $a$-maximization and $Z$-minimization,"
[arXiv:hep-th/0512240].


\bibitem{witten}
E. Witten, ``Anti-de Sitter space and holography," Adv. Theor. Math.
Phys. {\bf 2}, 253 (1998) [arXiv:hep-th/9802150].



\bibitem{trr}
E.~Barnes, E.~Gorbatov, K.~Intriligator, M.~Sudano and J.~Wright,
  ``The exact superconformal $R$-symmetry minimizes $\tau_{RR}$,''
  Nucl.\ Phys.\ B {\bf 730}, 210 (2005)
  [arXiv:hep-th/0507137].





\bibitem{gaugedsugra}
M. G\"unaydin, G. Sierra and P.K. Townsend, ``Gauging the $D=5$
Maxwell-Einstein supergravity theories: more on Jordan algebras",
Nucl. Phys. B {\bf 253}, 573 (1985);
\\
A. Chou, R. Kallosh, J. Rahmfeld, S.-J. Rey, M. Shmakova and W.K.
Wong, ``Critical Points and Phase Transitions in 5-D
compactifications of M-theory", Nucl.\ Phys.\ B {\bf 508}, 147
(1997) [arXiv:hep-th/9704142] \\
 A. Ceresole and G. Dall'Agata, ``General matter coupled
${\cal N}=2, D=5$ gauged supergravity", Nucl. Phys. B {\bf 585}, 143
(2000).



\bibitem{bhk}
D. Berenstein, C. P. Herzog and I. R. Klebanov, ``Baryon spectra and
AdS/CFT correspondence," JHEP {\bf 0206}, 047 (2002)
[arXiv:hep-th/0202150].



\bibitem{krvn}
H.~J.~Kim, L.~J.~Romans and P.~van Nieuwenhuizen,
  ``The Mass Spectrum Of Chiral $N = 2$, $D = 10$ Supergravity On $S^5$,''
  Phys.\ Rev.\ D {\bf 32}, 389 (1985).


\bibitem{lmsr}
S.~Lee, S.~Minwalla, M.~Rangamani and N.~Seiberg,
  ``Three-point functions of chiral operators in $D = 4$, $\CN = 4$ SYM at  large
  $N$,''
  Adv.\ Theor.\ Math.\ Phys.\  {\bf 2}, 697 (1998)
  [arXiv:hep-th/9806074].


\bibitem{af}
G.~Arutyunov and S.~Frolov,
  ``Some cubic couplings in type IIB supergravity on $AdS_5\times S^5$ and
  three-point functions in SYM$_4$ at large $N$,''
  Phys.\ Rev.\ D {\bf 61}, 064009 (2000)
  [arXiv:hep-th/9907085].

\bibitem{me}
S.~Lee,
  ``AdS$_5$/CFT$_4$ four-point functions of chiral primary operators: Cubic
  vertices,''
  Nucl.\ Phys.\ B {\bf 563}, 349 (1999)
  [arXiv:hep-th/9907108].


\bibitem{bpt}
  S.~Benvenuti, L.~A.~Pando~Zayas and Y.~Tachikawa,
  ``Triangle Anomalies from Einstein Manifolds,''
  [arXiv:hep-th/0601054].








\end{thebibliography}

\end{document}